\documentclass[12pt]{iopart}
\usepackage{iopams} 
\usepackage{mathbbol} 
\usepackage[colorlinks, linkcolor =blue]{hyperref} 
\usepackage{amsthm} 
\usepackage{mathrsfs} 
\usepackage{graphicx} 
\usepackage{esint}
\usepackage{color}
\usepackage{epsf}
\usepackage{subfigure}
\usepackage{epstopdf}
\usepackage{caption}
\usepackage{enumerate}
\usepackage{array}
\usepackage{bm}

\newcommand{\bol}[1]{\boldsymbol{#1}}
\newcommand{\Imag}{\mathrm{Im}}
\newcommand{\sinc}{\mathrm{sinc}}

\begin{document}


\title{Quantum backflow for two identical particles}

\author{Maximilien Barbier$^{1,*}$ and Arseni Goussev$^{2,3}$}

\address{$^1$Scottish Universities Physics Alliance, University of the West of Scotland, Paisley PA1 2BE, Scotland, United Kingdom \\
$^2$Section of Mathematics, University of Geneva, Rue du Conseil-Général 7-9, 1205 Genève, Switzerland\\
$^3$Quantum Physics Corner Ltd, 20-22 Wenlock Road, London N1 7GU, United Kingdom\\
$^*$ Author to whom any correspondence should be addressed.}

\ead{Maximilien.Barbier@uws.ac.uk}


\begin{abstract}
Quantum mechanics introduces the possibility for particles to move in a direction opposite to their momentum -- a counter-intuitive and classically impossible phenomenon known as quantum backflow. The magnitude of this effect is relatively small, making its experimental observation, which has yet to be achieved, particularly challenging. Here, we investigate the influence of quantum statistics on the maximal backflow attainable for two identical particles confined to a ring. Notably, we demonstrate that the fermionic statistics significantly impedes quantum backflow compared to the bosonic statistics. Our findings suggest that any future experimental realization of quantum backflow should prioritize systems involving bosons rather than fermions.
\end{abstract}

\vspace{2pc}
\noindent{\it Keywords}: quantum backflow, identical particles, quantum statistics, periodic boundary conditions


\section{Introduction}\label{intro}

In a striking departure from the principles of classical mechanics, quantum mechanics enables the appearance of interference effects. The latter give rise to a multitude of phenomena that challenge our classical intuition. For instance, a quantum electron can no longer be pictured as going through one slit or the other in the celebrated double-slit experiment, and its probability of detection behind the slits exhibits the characteristic interference fringes~\cite{Jon61,TEM89}. Another peculiar consequence of interference is that a quantum particle may seemingly move in the direction opposite to its momentum. In other words, in quantum mechanics, probability can flow counter to a quantum particle’s momentum -- a surprising phenomenon, forbidden in classical mechanics, known as quantum backflow.

Quantum backflow has been first identified as one of the difficulties that arises when one attempts at defining the notion of arrival times in quantum mechanics~\cite{All69,Kij74,Wer88}. It has then been investigated in its own right by Bracken and Melloy~\cite{BM94}, who showed in particular the existence of this effect for normalized states in the case of a free nonrelativistic quantum particle moving on a line with a positive momentum. Furthermore, they demonstrated that the maximal amount of probability that can flow in the negative direction, i.e. opposite to the momentum's direction, admits the nontrivial bound
\begin{eqnarray}
c_{\mathrm{line}} \approx 0.0384517 \, ,
\label{BM_bound}
\end{eqnarray}
which is now commonly referred to as the Bracken-Melloy bound. Although the exact value of $c_{\mathrm{line}}$ remains unknown\footnote{It has also been recently demonstrated in~\cite{TLN22} that $0.0315 \leqslant c_{\mathrm{line}} \leqslant 0.0725$.}, it has been numerically estimated using various techniques~\cite{BM94,EFV05,PGK06}. Surprisingly though, the Bracken-Melloy bound $c_{\mathrm{line}}$ can be shown~\cite{BM94} to be independent of Planck's constant. This hence immediately raises the question of the classical limit of quantum backflow (see Refs.~\cite{YHH12,Bar20,Bra21,MM22} for possible approaches in various contexts).

Following these pioneering investigations, quantum backflow has then been studied for a broad range of systems in various physical scenarios, such as a particle in the presence of a linear potential~\cite{MB98}, a relativistic particle~\cite{MB98,MB98_Dirac,ALS19}, a particle in a ring~\cite{Gou21,GQJ24}, two-dimensional~\cite{Str12,BGS23,PPR20} and many-particle~\cite{Bar20} scenarios, or decaying~\cite{DT19} and dissipative systems~\cite{MM22,MM20,MM20_Erratum,MM20_2part}. Many examples of backflowing states have been explored analytically~\cite{YHH12,HGL13,Str24,Chr24}. Furthermore, since quantum backflow is rooted in interference, phenomena akin to backflow also arise in other systems that support interference. For instance, backflow has also been predicted to occur for classical light~\cite{Ber10}.

While optical backflow has been observed recently~\cite{EZB20,DGG22,GDG23}, an experimental observation of quantum backflow still remains to be performed. Various approaches have been suggested to reach this milestone, such as using Bose-Einstein condensates~\cite{PTM13,MPM14}, devising an ``experiment-friendly'' formulation of quantum backflow~\cite{MYD21,BG21} or measuring times of arrival~\cite{BBM24}. Therefore, it is essential to broaden the range of physical platforms that hold promise for the prospective observation of this still elusive quantum phenomenon.

The goal of this work is to investigate quantum backflow in systems of two identical particles to understand the influence of quantum statistics. Many-particle quantum backflow remains a largely unexplored field\footnote{To date, many-particle backflow has been studied only in two cases: $N$ particles on a line~\cite{Bar20}, and two particles in dissipative systems~\cite{MM20_2part}.}, particularly relevant for prospective experimental observations. Indeed, the latter are likely to be conducted in many-particle systems (e.g., Bose-Einstein condensates, as suggested in~\cite{PTM13,MPM14}) rather than with single particles. Gaining a detailed understanding of quantum backflow in two-particle systems represents a logical first step in this direction.

For our analysis, we focus on the scenario of two particles moving freely in a circular ring rather than along an infinite line. This choice is motivated by two key factors. First, the maximum quantum backflow magnitude for a single particle is significantly larger in a ring than on a line. Indeed, numerical results in Ref.~\cite{Gou21} show that the maximal backflow achievable in a ring is given by
\begin{eqnarray}
c_{\mathrm{ring}} \approx 0.116816 \, ,
\label{c_ring_1_part}
\end{eqnarray}
which exceeds the Bracken-Melloy bound~\eref{BM_bound} by more than threefold. Moreover, the backflow-maximizing state in the ring, as identified in~\cite{Gou21}, is far less singular than its counterpart on the line~\cite{BM94}, making it potentially much more feasible for experimental implementation. Second, the discrete energy spectrum of particle motion in a ring renders this system more amenable to both analytical and numerical investigation.

The key finding of our analysis is that the quantum statistics -- whether bosonic or fermionic -- of two particles in a ring critically influences the maximally achievable backflow. Specifically, we analytically show that the maximal backflow for two bosons is exactly twice that of the maximal backflow for a single particle. In contrast, our numerical simulations reveal that the maximal backflow for two fermions is significantly smaller than the corresponding single-particle value. These results suggest the following practical conclusion: any prospective experimental realization of quantum backflow should prioritize systems involving bosons rather than fermions.

The paper is structured as follows. We discuss in section~\ref{QB_ring_sec} how quantum backflow can be formulated in a ring, first recalling the single-particle scenario and then addressing the case of two identical particles. We then compute in section~\ref{min_sec} the maximal amount of backflow that is achievable for two identical particles in a ring, separately treating the case of bosons and of fermions. Finally, a summary of our findings, as well as concluding remarks, are presented in section~\ref{ccl}.


\section{Quantum backflow in a ring}\label{QB_ring_sec}

We first recall in section~\ref{one_part_subsec} how quantum backflow is formulated for a single particle moving in a circular ring. We then discuss in section~\ref{two_part_subsec} how this problem can be generalized to a two-particle scenario.


\subsection{Single-particle quantum backflow in a ring}\label{one_part_subsec}

Here we summarize how quantum backflow has been formulated in~\cite{Gou21} for a single nonrelativistic particle of mass $\mu$ in a circular ring of radius $R$. We assume that the particle moves freely in the ring, the latter being assumed to lie in the Cartesian $(x,y)$ plane. The state of the particle at time $t$ is described by the wave function $\psi(\theta, t)$, where $\theta$ denotes the polar angle. The wave function is normalized, that is
\begin{eqnarray}
\int_{0}^{2 \pi} d \theta \, | \psi (\theta, t) |^2 = 1 \, ,
\label{normal_psi_1_part}
\end{eqnarray}
and periodic, that is
\begin{eqnarray}
\psi(\theta + 2 \pi, t) = \psi(\theta,t) \, .
\end{eqnarray}
Note in particular that the wave function $\psi(\theta, t)$ is dimensionless. It obeys the time-dependent Schr\"odinger equation
\begin{eqnarray}
i \hbar \frac{\partial \psi}{\partial t} = H \psi \, ,
\label{TDSE_ring_1_part}
\end{eqnarray}
where the Hamiltonian $H$ is
\begin{eqnarray}
H = - \frac{\hbar^2}{2 \mu R^2} \frac{\partial^2}{\partial \theta^2} \, .
\label{H_ring_1_part}
\end{eqnarray}

The eigenenergies $E_m$ and the energy eigenstates $\psi_m (\theta)$, i.e. the solutions of the time-independent Schr\"odinger equation
\begin{eqnarray}
H \psi_m (\theta) = E_m \psi_m (\theta) \, ,
\label{TISE_1_part}
\end{eqnarray}
are given by
\begin{eqnarray}
E_m = \frac{\hbar^2 m^2}{2 \mu R^2} \qquad \mbox{and} \qquad \psi_m (\theta) = \frac{1}{\sqrt{2 \pi}} \, e^{i m \theta} \quad , \quad m \in \mathbb{Z} \, .
\label{E_m_psi_m_1_part}
\end{eqnarray}
The set $\{ \psi_m \}_{m \in \mathbb{Z}}$ is orthonormal, that is
\begin{eqnarray}
\int_{0}^{2 \pi} d\theta \, \psi_{m}^*(\theta) 
\, \psi_{m'}(\theta) = \delta_{mm'}  \, ,
\label{ON_set_1_part}
\end{eqnarray}
and complete, providing an orthonormal basis of the Hilbert space $\mathcal{H}_{\mathrm{ring}}$ associated with the particle.

The canonical angular-momentum operator $\bol{L}$ is
\begin{eqnarray}
\bol{L} = \bol{r} \times \bol{p} \, ,
\label{L_def}
\end{eqnarray}
where $\bol{p} = - i \hbar \bol{\nabla}$ denotes the momentum operator. Since the ring is confined in the $(x,y)$ plane, the angular momentum points normally to the plane:
\begin{eqnarray}
\bol{L} = L_z \bol{e}_z = - i \hbar \frac{\partial}{\partial \theta} \bol{e}_z \, ,
\label{L_vert}
\end{eqnarray}
where $\bol{e}_z $ denotes the unit Cartesian vector along the $z$ axis. The energy eigenstates~\eref{E_m_psi_m_1_part} are also eigenstates of $L_z$, and we have
\begin{eqnarray}
L_z \psi_m (\theta) = m \hbar \psi_m (\theta) \, .
\label{L_z_psi_m}
\end{eqnarray}
The angular momentum is thus non-negative for any $m \geqslant 0$. In the study of quantum backflow in a ring~\cite{Gou21}, attention is restricted to quantum states with non-negative angular momentum:
\begin{eqnarray}
\psi(\theta,t) = \sum_{m=0}^{\infty} c_m \psi_m (\theta) e^{-i E_m t / \hbar} \, .
\label{non_neg_superp_1_part}
\end{eqnarray}
For a state of the form~\eref{non_neg_superp_1_part}, quantum backflow occurs whenever the probability current
\begin{eqnarray}
j(\theta,t) = \frac{\hbar}{\mu R^2} \Imag \left( \psi^* \frac{\partial \psi}{\partial \theta} \right)
\label{current_def_1_part}
\end{eqnarray}
takes on a negative value, $j(\theta,t) < 0$, for some $\theta$ and $t$. For later reference, we point out that the normalization condition~\eref{normal_psi_1_part} applied to a state of the form~\eref{non_neg_superp_1_part} reads
\begin{eqnarray}
\sum_{m=0}^{\infty} |c_m|^2 = 1 \, ,
\label{normal_superp_1_part}
\end{eqnarray}
while the substitution of~\eref{non_neg_superp_1_part} into~\eref{current_def_1_part} yields for the current
\begin{eqnarray}
j(\theta,t) = \frac{\hbar}{2 \mu R^2} \sum_{m,n=0}^{\infty} (m+n) c_m^* c_n \psi_m^* (\theta) \psi_n (\theta) e^{i(E_m-E_n)t/\hbar} \, .
\label{current_superp_1_part}
\end{eqnarray}

While a negative sign of the current~\eref{current_superp_1_part} unambiguously signals the occurrence of quantum backflow, its magnitude hardly provides a relevant quantifier of the effect. Indeed, the analysis performed in~\cite{GQJ24} shows that $j(\theta,t)$ can take on arbitrarily small negative values. Furthermore, in practice, a measurement of the current would typically be performed over a finite time interval. Therefore, a more relevant quantifier of the magnitude of quantum backflow is the time-integrated current $\Delta_1$ at a given point of the ring, say $\theta = 0$, over a time interval $T>0$ :
\begin{eqnarray}
\Delta_1 = \int_{-T/2}^{T/2} dt \, j(0,t) \, .
\label{Delta_1_part}
\end{eqnarray}
The substitution of~\eref{current_superp_1_part} into~\eref{Delta_1_part} yields~\cite{Gou21}
\begin{eqnarray}
\Delta_1 = \sum_{m,n=0}^{\infty} c_m^* K_{mn} c_n \, ,
\label{Delta_expr_1_part}
\end{eqnarray}
where the kernel $K_{mn}$ is given by
\begin{eqnarray}
K_{mn} = \frac{\alpha}{\pi} (m+n) \sinc \left[ \alpha (m^2 - n^2) \right] \, ,
\label{K_mn_1_part}
\end{eqnarray}
where
\begin{eqnarray}
\alpha = \frac{\hbar T}{4 \mu R^2}
\label{alpha_def}
\end{eqnarray}
and $\sinc z = \sin (z) / z$. Note that, since the function $\sinc \, z$ is even, the kernel $K_{mn}$ is symmetric, that is
\begin{eqnarray}
K_{nm} = K_{mn} \, .
\label{K_mn_sym}
\end{eqnarray}
As shown numerically in~\cite{Gou21}, the minimization of $\Delta_1$ over the space of non-negative-angular-momentum states of the form~\eref{non_neg_superp_1_part} under the normalization constraint~\eref{normal_superp_1_part} yields the following lower bound:
\begin{eqnarray}
\inf\limits_{\psi} \Delta_1 = - c_{\mathrm{ring}} \, ,
\label{max_QB_1_part}
\end{eqnarray}
where $c_{\mathrm{ring}} \approx 0.116816$ is the constant given by~\eref{c_ring_1_part}.

It is worth mentioning that the analysis performed in~\cite{Gou21} also considers the presence of a uniform magnetic field perpendicular to the ring, i.e. along $+\bol{e}_z$. As was shown in~\cite{Gou21}, a magnetic field actually reduces the maximally achievable magnitude of the backflow effect. In other words, the presence of a magnetic field appears to be detrimental to the occurrence of quantum backflow for a charged particle in a ring. A similar observation can be made in the case of a particle moving on a line in the presence of a linear potential~\cite{MB98}.

We now discuss how the problem of quantum backflow in a ring can be generalized to a two-particle scenario.


\subsection{Two-particle quantum backflow in a ring}\label{two_part_subsec}

We now assume that we have two identical particles of mass $\mu$, moving in the same circular ring of radius $R$ as in section~\ref{one_part_subsec}. We denote by $\theta_1$ the polar angle associated with one particle, and by $\theta_2$ the polar angle associated with the other particle. The state of the system at time $t$ is now described by a two-particle wave function $\Psi(\theta_1, \theta_2, t)$, which is assumed to be normalized,
\begin{eqnarray}
\int_{0}^{2 \pi} d \theta_1 \int_{0}^{2 \pi} d \theta_2 \, |\Psi (\theta_1, \theta_2, t)|^2 = 1 \, ,
\label{normal_psi_2_part}
\end{eqnarray}
and periodic, namely here
\begin{eqnarray}
\Psi (\theta_1 + 2 \pi, \theta_2, t) = \Psi (\theta_1, \theta_2 + 2 \pi, t) = \Psi (\theta_1, \theta_2, t) \, .
\end{eqnarray}
As before, the wave function $\Psi(\theta_1, \theta_2, t)$ is dimensionless. Since the particles are identical, $\Psi$ must, in addition to the normalization constraint~\eref{normal_psi_2_part}, satisfy the symmetrization constraint, that is
\begin{eqnarray}
\Psi (\theta_2, \theta_1, t) = \sigma \Psi (\theta_1, \theta_2, t) \, ,
\label{sym_postulate_Psi}
\end{eqnarray}
where
\begin{eqnarray}
\sigma = \left\{\begin{array}{ll}
+1 \quad  &\mbox{for identical bosons} \\[0.2cm]
-1 \quad &\mbox{for identical fermions}
\end{array}\right. \, .
\label{sigma_def}
\end{eqnarray}

We assume that the two particles move freely along the ring, so that the wave function $\Psi$ obeys the free time-dependent Schr\"odinger equation
\begin{eqnarray}
i \hbar \frac{\partial \Psi}{\partial t} = - \frac{\hbar^2}{2 \mu R^2} \left( \frac{\partial^2}{\partial \theta_1^2} + \frac{\partial^2}{\partial \theta_2^2} \right) \Psi \, .
\label{TDSE_ring_2_part}
\end{eqnarray}
We emphasize that the free-particle description remains a physically meaningful and fruitful model, even when the two particles are electrically charged. This approach can be regarded as a specific case of the free-electron-gas model, which is routinely employed in condensed matter physics (see, e.g.,~\cite{AshMer}).

The Hilbert space $\mathcal{H}_2$ associated with this two-particle system is thus a tensor product of the single-particle Hilbert space $\mathcal{H}_{\mathrm{ring}}$, that is $\mathcal{H}_2 = \mathcal{H}_{\mathrm{ring}} \otimes \mathcal{H}_{\mathrm{ring}}$. Since the single-particle energy eigenstates $\psi_m (\theta)$, given by~\eref{E_m_psi_m_1_part}, provide an orthonormal basis of $\mathcal{H}_{\mathrm{ring}}$, the set $\{ \psi_{m_1} (\theta_1) \psi_{m_2} (\theta_2) \}_{m_1,m_2 \in \mathbb{Z}}$ provides an orthonormal basis\footnote{Note however that these two-particle basis states do \textit{not} individually satisfy the symmetrization constraint~\eref{sym_postulate_Psi}.} of $\mathcal{H}_2$. Therefore, in view of formulating quantum backflow for this two-particle system, here we consider states $\Psi(\theta_1, \theta_2, t)$ of the form
\begin{eqnarray}
\Psi (\theta_1, \theta_2, t) = \sum_{m_1,m_2=0}^{\infty} c_{m_1 m_2} \psi_{m_1} (\theta_1) \psi_{m_2} (\theta_2) e^{-i (E_{m_1} + E_{m_2}) t / \hbar} \, .
\label{non_neg_superp_2_part}
\end{eqnarray}
That is, we only consider two-particle states that are superpositions of single-particle states that have a non-negative angular momentum. The normalization condition~\eref{normal_psi_2_part} applied to the state~\eref{non_neg_superp_2_part} yields
\begin{eqnarray}
\sum_{m_1, m_2=0}^{\infty} |c_{m_1 m_2}|^2 = 1 \, ,
\label{normal_superp_2_part}
\end{eqnarray}
whereas the symmetrization constraint~\eref{sym_postulate_Psi} applied to the state~\eref{non_neg_superp_2_part} yields
\begin{eqnarray}
c_{m_2 m_1} = \sigma c_{m_1 m_2} \qquad , \qquad \forall m_1, m_2 \geqslant 0 \, .
\label{sym_postulate_c_coef}
\end{eqnarray}

To quantify quantum backflow in our two-particle system, we focus on the particle-number current\footnote{We note that the formulation of many-particle quantum backflow proposed in~\cite{Bar20} cannot be applied in the present case. Indeed,~\cite{Bar20} deals with a many-particle system moving freely on an infinite line, which hence allows to define probabilities of finding a particle at a negative position or at a positive position. Such probabilities cannot be defined for a periodic system such as a ring.} (see e.g.~\cite{Gottfried})
\begin{eqnarray}
\fl J (\theta, t) = \frac{\hbar}{\mu R^2} \Imag \left\{ \int_{0}^{2 \pi} d \varphi \left[ \Psi^*(\theta, \varphi, t) \frac{\partial \Psi(\theta, \varphi, t)}{\partial \theta} + \Psi^*(\varphi, \theta, t) \frac{\partial \Psi(\varphi, \theta, t)}{\partial \theta} \right] \right\} \, .
\label{J_def_2_part}
\end{eqnarray}
This current represents the number of particles crossing the point $\theta$ at time $t$: if $J > 0$ the flow is in the positive (counter-clockwise) direction, whereas if $J < 0$ the flow is in the negative (clockwise) direction. If the particles that we consider have an electric charge, $J$, multiplied by the charge, represents the electric current at the point $\theta$ at time $t$. This current is e.g. widely used in chemistry as the electron current density~\cite{Seb21}. The quantity~\eref{J_def_2_part} is thus the two-particle-in-a-ring analog of the single-particle current~\eref{current_def_1_part}. It is the current associated with the particle-number density $\rho(\theta,t)$ defined by (see e.g.~\cite{Gottfried})
\begin{eqnarray}
\rho(\theta,t) = \int_{0}^{2 \pi} d \varphi \left[ |\Psi (\theta, \varphi, t)|^2 + |\Psi (\varphi, \theta, t)|^2 \right] \, .
\label{rho_def}
\end{eqnarray}
The density~\eref{rho_def} integrated over the whole ring yields the total number of particles,
\begin{eqnarray}
\int_{0}^{2 \pi} d\theta \, \rho(\theta,t) = 2 \, ,
\label{int_dens}
\end{eqnarray}
at any $t$. The current~\eref{J_def_2_part} and the density~\eref{rho_def} satisfy the continuity equation~\cite{Gottfried}
\begin{eqnarray}
\frac{\partial \rho}{\partial t} + \frac{\partial J}{\partial \theta} = 0 \, .
\label{cont_eq_2_part}
\end{eqnarray}

Quantum backflow occurs whenever the current $J$ takes on a negative value, i.e. $J (\theta, t) < 0$ for some $\theta$ and $t$, for a state of the form~\eref{non_neg_superp_2_part}. The underlying intuition can be illustrated by calculating the current $J$ for a single basis state $\Psi_{m_1 m_2}$, which is of the form
\begin{eqnarray}
\Psi_{m_1 m_2} (\theta_1, \theta_2, t) = \psi_{m_1} (\theta_1) \psi_{m_2} (\theta_2) e^{-i (E_{m_1} + E_{m_2}) t / \hbar} \, .
\label{single_basis_state_2_part}
\end{eqnarray}
Substituting~\eref{single_basis_state_2_part} into~\eref{J_def_2_part} yields
\begin{eqnarray}
\left. J (\theta, t) \right|_{\Psi = \Psi_{m_1 m_2}} = (m_1 + m_2) \frac{\hbar}{2 \pi \mu R^2} \, ,
\label{current_single_basis_state}
\end{eqnarray}
which is thus clearly non-negative for any $m_1,m_2 \geqslant 0$. This shows that a state of the form~\eref{non_neg_superp_2_part} consists of a superposition of eigenstates that each individually exhibits a non-negative current $J$. Quantum backflow hence occurs when such a superposition yields a negative current $J$, for some angle $\theta$ on the ring and time $t$.

We now use the fact that the two particles are identical to rewrite $J (\theta, t)$ and $\rho (\theta, t)$ in a more compact form. Combining~\eref{J_def_2_part} and~\eref{rho_def} with the symmetrization constraint~\eref{sym_postulate_Psi} yields, also using the fact that $|\sigma| = 1$ in view of~\eref{sigma_def},
\begin{eqnarray}
J (\theta, t) = \frac{2 \hbar}{\mu R^2} \Imag \left[ \int_{0}^{2 \pi} d \varphi \, \Psi^*(\theta, \varphi, t) \frac{\partial \Psi(\theta, \varphi, t)}{\partial \theta} \right]
\label{compact_J_def_2_part}
\end{eqnarray}
and
\begin{eqnarray}
\rho(\theta,t) = 2 \int_{0}^{2 \pi} d \varphi \, |\Psi (\theta, \varphi, t)|^2 \, .
\label{compact_rho_def}
\end{eqnarray}

Now, we construct the two-particle analog $\Delta_2$ of the single-particle time-integrated current $\Delta_1$ defined in~\eref{Delta_1_part} by replacing the single-particle current $j$ by the two-particle current $J$ in~\eref{Delta_1_part}, so that we define
\begin{eqnarray}
\Delta_2 = \int_{-T/2}^{T/2} dt \, J (0, t) \, .
\label{Delta_2_def}
\end{eqnarray}
Given that $J$ can be interpreted as the electric current in the case of charged particles, we will henceforth refer to $\Delta_2$ as the charge transfer. Indeed, if the particles are electrically charged, $\Delta_2$ represents the total electric charge passing through the point $\theta = 0$ during the time interval $T$. Note that $\Delta_2$ can be larger than the total number of particles present in the ring, since a same particle can cross the point $\theta = 0$ more than once during the time interval $T$. Substituting the expression~\eref{non_neg_superp_2_part} of $\Psi$ into the expression~\eref{compact_J_def_2_part} for $J$ and setting $\theta = 0$, we obtain
\begin{eqnarray}
J(0,t) = \frac{\hbar}{2 \pi \mu R^2} \sum_{m,n,k=0}^{\infty} (m+n) c_{m k}^* c_{n k} e^{i (E_{m} - E_{n}) t / \hbar} \, .
\label{J_theta_zero}
\end{eqnarray}
Finally, substituting~\eref{J_theta_zero} into~\eref{Delta_2_def} and performing the integral, we find the following expression for the charge transfer:
\begin{eqnarray}
\Delta_2 = 2 \sum_{m,n,k=0}^{\infty} c_{mk}^* K_{mn} c_{nk} \, ,
\label{Delta_2_expr}
\end{eqnarray}
where $K_{mn}$ is nothing but the single-particle kernel~\eref{K_mn_1_part}. Note the factor of 2 in the right-hand side of~\eref{Delta_2_expr}, which simply accounts for the fact that we have two identical particles here.

To identify the backflow-maximizing configuration -- corresponding to the smallest possible (negative) charge transfer -- we proceed to minimizing $\Delta_2$ while enforcing both the normalization and symmetrization constraints, addressing the bosonic and fermionic cases separately.


\section{Minimization of the charge transfer \texorpdfstring{$\Delta_{2}$}{Delta}}\label{min_sec}

The general constrained minimization problem we aim to solve here involves minimizing the charge transfer~\eref{Delta_2_expr} with respect to the complex coefficients $c_{mn}$, subject to the normalization constraint~\eref{normal_superp_2_part} and the symmetrization constraint~\eref{sym_postulate_c_coef}. Without loss of generality, the coefficients $c_{mn}$ can be taken as real (see~\ref{real_coef_app} for a proof). That is, the above-mentioned constrained minimization problem is strictly equivalent to minimizing the quantity
\begin{eqnarray}
\Delta_2 = 2 \sum_{m,n,k=0}^{\infty} c_{mk} K_{mn} c_{nk}
\label{real_Delta_2_gen_expr}
\end{eqnarray}
with respect to the \textit{real} coefficients $c_{mn}$, with the two constraints~\eref{normal_superp_2_part} and~\eref{sym_postulate_c_coef} hence reading
\begin{eqnarray}
\sum_{m,n=0}^{\infty} c_{m n}^2 = 1
\label{real_norm_constraint_def}
\end{eqnarray}
and
\begin{eqnarray}
c_{n m} = \sigma c_{m n} \qquad , \qquad \forall m,n \geqslant 0 \, .
\label{real_sym_constraint_def}
\end{eqnarray}

Our strategy for minimizing the charge transfer~\eref{real_Delta_2_gen_expr} is to introduce an integer parameter $N \geqslant 1$ and consider the quantity $\Delta_2^{(N)}$ defined by
\begin{eqnarray}
\Delta_2^{(N)} = 2 \sum_{k=0}^N \sum_{m=0}^N \sum_{n=0}^N c_{mk} K_{mn} c_{nk} \, .
\label{Delta2_finite_N}
\end{eqnarray}
Comparing~\eref{real_Delta_2_gen_expr} and~\eref{Delta2_finite_N}, we readily see that
\begin{eqnarray}
\Delta_2 = \lim\limits_{N \to \infty} \Delta_2^{(N)} \, .
\label{Delta_2_Delta_2_N_rel}
\end{eqnarray}
The idea is thus to find the minimal value of $\Delta_2^{(N)}$ for a given value of $N$, and to then evaluate the limit of this minimal value as $N \to \infty$. The quantity $\Delta_2^{(N)}$ must of course be minimized under the normalization constraint, which here reads
\begin{eqnarray}
\sum_{k=0}^N \sum_{m=0}^N c_{mk}^2 = 1 \, ,
\label{normalization_finite_N}
\end{eqnarray}
as well as the symmetrization constraint,
\begin{eqnarray}
c_{mk} = \sigma c_{km} \qquad , \qquad \forall \quad 0 \leqslant k,m \leqslant N \, .
\label{sym_constraint_finite_N}
\end{eqnarray}

We hence have two distinct physical problems to treat depending on the nature of the particles, i.e. depending on whether they are bosons or fermions:

\begin{enumerate}
	\item {\it Bosons:} The minimal charge transfer $\mathcal{Q}_{\mathrm{B}}^{(N)}$ for two identical bosons is obtained by minimizing $\Delta_2^{(N)}$, given by Eq.~\eref{Delta2_finite_N}, subject to the normalization constraint~\eref{normalization_finite_N} and the symmetrization constraint~\eref{sym_constraint_finite_N} for $\sigma = +1$, i.e.
	\begin{eqnarray}
		c_{mk} = c_{km} \qquad , \qquad \forall \quad 0 \leqslant k,m \leqslant N \, .
	\label{symmetry_bosons}
	\end{eqnarray}
	
	\item {\it Fermions:} The minimal charge transfer $\mathcal{Q}_{\mathrm{F}}^{(N)}$ for two identical fermions is obtained by minimizing $\Delta_2^{(N)}$, given by Eq.~\eref{Delta2_finite_N}, subject to the normalization constraint~\eref{normalization_finite_N} and the symmetrization constraint~\eref{sym_constraint_finite_N} for $\sigma = -1$, i.e.
	\begin{eqnarray}
		c_{mk} = -c_{km} \qquad , \qquad \forall \quad 0 \leqslant k,m \leqslant N \, .
	\label{symmetry_fermions}
	\end{eqnarray}
\end{enumerate}

\noindent In both cases, we then need to evaluate the limit of $\mathcal{Q}_{\mathrm{B}}^{(N)}$ (for bosons) and $\mathcal{Q}_{\mathrm{F}}^{(N)}$ (for fermions) as $N \to \infty$. We first treat the problem of bosons in section~\ref{id_bosons_subsec}, and then we treat the problem of fermions in section~\ref{id_fermions_subsec}.


\subsection{Bosons}\label{id_bosons_subsec}

Here we determine the minimal charge transfer $\mathcal{Q}_{\mathrm{B}}$ for two identical bosons. We proceed in three steps:

\begin{enumerate}
\item We first determine in section~\ref{bos_norm_const_subsubsec} the minimal value $\mathcal{Q}_{\mathrm{D}}^{(N)}$ of $\Delta_2^{(N)}$ subject to the normalization constraint~\eref{normalization_finite_N} only;

\item From our expression of $\mathcal{Q}_{\mathrm{D}}^{(N)}$, we then determine in section~\ref{bos_sym_const_subsubsec} the minimal value $\mathcal{Q}_{\mathrm{B}}^{(N)}$ of $\Delta_2^{(N)}$ subject to both the normalization constraint~\eref{normalization_finite_N} and the symmetrization constraint~\eref{symmetry_bosons};

\item Finally, we find in section~\ref{bos_N_infty_subsubsec} the minimal charge transfer $\mathcal{Q}_{\mathrm{B}}$ by taking the limit of our expression of $\mathcal{Q}_{\mathrm{B}}^{(N)}$ as $N \to \infty$.
\end{enumerate}


\subsubsection{Minimal charge transfer under the normalization constraint}\label{bos_norm_const_subsubsec}

Here we determine the minimal value $\mathcal{Q}_{\mathrm{D}}^{(N)}$ of $\Delta_2^{(N)}$ subject to the normalization constraint~\eref{normalization_finite_N} only.

To this end, we first rewrite $\Delta_2^{(N)}$, as given by~\eref{Delta2_finite_N}, in the form
\begin{eqnarray}
\Delta_2^{(N)} = \sum_{k=0}^N \Delta_{2,k}^{(N)} \, ,
\label{Delta_2_Delta_2_k_rel}
\end{eqnarray}
with
\begin{eqnarray}
\Delta_{2,k}^{(N)} = 2 \sum_{m=0}^N \sum_{n=0}^N c_{mk} K_{mn} c_{nk} \, ,
\label{Delta2_k-th_component}
\end{eqnarray}
and we rewrite the normalization constraint~\eref{normalization_finite_N} as the following set of conditions:
\begin{eqnarray}
\sum_{m=0}^N c_{mk}^2 = \mu_k \, ,
\label{normalization_k_th_component}
\end{eqnarray}
for $k = 0, \ldots, N$, where $\mu_k \geqslant 0$ are such that
\begin{eqnarray}
\sum_{k=0}^N \mu_k = 1 \, .
\label{sum_of_mu_k}
\end{eqnarray}
Thus, the problem of minimizing $\Delta_2^{(N)}$ under the normalization constraint~\eref{normalization_finite_N} separates into $N+1$ independent minimization problems, where the $k^{\mathrm{th}}$ problem is concerned with minimizing $\Delta_{2,k}^{(N)}$, given by~\eref{Delta2_k-th_component}, subject to the constraint~\eref{normalization_k_th_component}. Equation~\eref{sum_of_mu_k} is then used to combine the sets of the expansion coefficients $\{ c_{0k}, \ldots, c_{Nk} \}$, obtained in each of the $N+1$ minimization problems, into a single normalized set $\{ c_{00}, \ldots, c_{NN} \}$ of $(N+1)^2$ expansion coefficients minimizing $\Delta_2^{(N)}$.

The minimization of $\Delta_{2,k}^{(N)}$ with respect to the set of $N+1$ coefficients $\{ c_{0k}, \ldots, c_{Nk} \}$, subject to the constraint~\eref{normalization_k_th_component}, is done in the usual manner. Introducing a Lagrange multiplier $2 \lambda$, one performs the unconstrained minimization of $\Delta_{2,k}^{(N)} - 2 \lambda \sum_{m=0}^N c_{mk}^2$, which gives rise to the following Euler-Lagrange equation:
\begin{eqnarray}
	\sum_{n=0}^N K_{mn} c_{nk} = \lambda c_{mk} \, .
\label{EL_k-th_component}
\end{eqnarray}
This equation is nothing but the Euler-Lagrange equation obtained in the case of a single particle in a ring~\cite{Gou21}. The smallest eigenvalue $\lambda$ is thus
\begin{eqnarray}
	\lambda_{\mathrm{ring}}^{(N)} = \min \lambda \, ,
\label{min_lambda}
\end{eqnarray}
and the corresponding eigenvector is
\begin{eqnarray}
	c_{mk} = \nu_k \tilde{c}_m \, .
\label{min_c_mk}
\end{eqnarray}
Here, $\{ \tilde{c}_0, \ldots, \tilde{c}_N \}$ is the set of expansion coefficients minimizing the probability transfer in the single-particle case, which is taken to be normalized to unity,
\begin{eqnarray}
	\sum_{m=0}^N \tilde{c}_m^2 = 1 \, .
\label{normalization_c_m}
\end{eqnarray}
Both $\lambda_{\mathrm{ring}}^{(N)}$ and $\tilde{c}_m$ have been numerically determined in~\cite{Gou21}. The proportionality constant $\nu_k$ in~\eref{min_c_mk} is then chosen to fulfill the normalization constraint~\eref{normalization_k_th_component}:
\begin{eqnarray}
	\mu_k = \sum_{m=0}^N c_{mk}^2 = \sum_{m=0}^N \tilde{c}_m^2 \nu_k^2 = \nu_k^2 \, ,
\end{eqnarray}
or
\begin{eqnarray}
	\nu_k = \sqrt{\mu_k} \, .
\label{nu_k}
\end{eqnarray}
In view of Eqs.~\eref{Delta2_k-th_component}, \eref{EL_k-th_component}, \eref{min_lambda}, \eref{min_c_mk}, and \eref{normalization_c_m}, we have
\begin{eqnarray}
	\min \Delta_{2,k}^{(N)} &= 2 \sum_{m=0}^N \nu_k \tilde{c}_m \sum_{n=0}^N K_{mn} \nu_k \tilde{c}_n \nonumber \\
	&= 2 \sum_{m=0}^N \nu_k \tilde{c}_m \lambda_{\mathrm{ring}}^{(N)} \nu_k \tilde{c}_m \nonumber \\
	&= 2 \nu_k^2 \lambda_{\mathrm{ring}}^{(N)} \, ,
\end{eqnarray}
and, finally, using Eq.~\eref{nu_k},
\begin{eqnarray}
	\min \Delta_{2,k}^{(N)} = 2 \mu_k \lambda_{\mathrm{ring}}^{(N)} \, .
\end{eqnarray}

Therefore, using the relation~\eref{Delta_2_Delta_2_k_rel}, we find that the minimal value $\mathcal{Q}_{\mathrm{D}}^{(N)}$ of the charge transfer $\Delta_{2}^{(N)}$ under the normalization constraint~\eref{normalization_finite_N} only is given by
\begin{eqnarray}
	\mathcal{Q}_{\mathrm{D}}^{(N)} = \sum_{k=0}^N \min \Delta_{2,k}^{(N)} = 2 \lambda_{\mathrm{ring}}^{(N)} \sum_{k=0}^N \mu_k \, ,
\end{eqnarray}
that, in view of~\eref{sum_of_mu_k}, is
\begin{eqnarray}
	\mathcal{Q}_{\mathrm{D}}^{(N)} = 2 \lambda_{\mathrm{ring}}^{(N)} \,.
\label{Q_D}
\end{eqnarray}
The expansion coefficients of the state corresponding to the minimal charge transfer are given by
\begin{eqnarray}
	c_{mk} = \tilde{c}_m \sqrt{\mu_k} \,.
\label{c_mk_for_distinguishable_particles}
\end{eqnarray}
Here, $\mu_0, \ldots, \mu_N$ are any non-negative numbers satisfying Eq.~\eref{sum_of_mu_k}. Clearly, there are infinitely many ways to choose these numbers, meaning that the backflow-maximizing state is infinitely degenerate \textit{when only the normalization constraint is taken into account}.

We now determine the minimal value of $\Delta_{2}^{(N)}$ under \textit{both} the normalization constraint \textit{and} the symmetrization constraint.


\subsubsection{Minimal charge transfer under the normalization and symmetrization constraints}\label{bos_sym_const_subsubsec}

We just determined in section~\ref{bos_norm_const_subsubsec} the minimal value $\mathcal{Q}_{\mathrm{D}}^{(N)}$ of $\Delta_2^{(N)}$ subject to the normalization constraint~\eref{normalization_finite_N} only. We now build up on this result to find the minimal value $\mathcal{Q}_{\mathrm{B}}^{(N)}$ of $\Delta_2^{(N)}$ subject to \textit{both} the normalization constraint~\eref{normalization_finite_N} \textit{and} the symmetrization constraint~\eref{symmetry_bosons}.

To this end, we first notice that
\begin{eqnarray}
	\mathcal{Q}_{\mathrm{B}}^{(N)} \geqslant \mathcal{Q}_{\mathrm{D}}^{(N)} \, ,
\end{eqnarray}
as we now have an additional constraint [namely the symmetrization one~\eref{symmetry_bosons}] as compared to the minimization problem leading to $\mathcal{Q}_{\mathrm{D}}^{(N)}$. The equality sign is achieved if and only if there exists a set of expansion coefficients $\{ c_{00}, \ldots, c_{NN} \}$ that i) minimizes $\Delta_2^{(N)}$ under the normalization constraint and ii) also satisfies the symmetrization constraint~\eref{symmetry_bosons}. As we now show, such a set indeed exists.

The expansion coefficients minimizing $\Delta_2^{(N)}$ under the normalization constraint are those that satisfy Eqs.~\eref{c_mk_for_distinguishable_particles} and \eref{sum_of_mu_k}. Let us choose
\begin{eqnarray}
	\mu_k = \tilde{c}_k^2 \, ,
\end{eqnarray}
so that
\begin{eqnarray}
	c_{mk} = \tilde{c}_m \tilde{c}_k \, .
\label{c_mk_for_bosons}
\end{eqnarray}
This choice clearly fulfills both the symmetrization constraint, Eq.~\eref{symmetry_bosons}, and the normalization constraint, Eq.~\eref{sum_of_mu_k} [in view of~Eq.~\eref{normalization_c_m}].

Therefore, we have
\begin{eqnarray}
	\mathcal{Q}_{\mathrm{B}}^{(N)} = \mathcal{Q}_{\mathrm{D}}^{(N)} \, ,
	\label{Q_B_eq_Q_D}
\end{eqnarray}
and thus in view of the expression~\eref{Q_D} of $\mathcal{Q}_{\mathrm{D}}^{(N)}$ we find that
\begin{eqnarray}
	\mathcal{Q}_{\mathrm{B}}^{(N)} = 2 \lambda_{\mathrm{ring}}^{(N)} \, .
	\label{Q_B_N_final_expr}
\end{eqnarray}
However, unlike in the case analysed in section~\ref{bos_norm_const_subsubsec}, where only the normalization constraint was considered, the backflow-maximizing state becomes non-degenerate \textit{when both the normalization and symmetrization constraints are taken into account}. Specifically, this state is now the product state described by Eq.~\eref{c_mk_for_bosons}.


\subsubsection{Limit as \texorpdfstring{$N \to \infty$}{N going to infinity}}\label{bos_N_infty_subsubsec}

It is now straightforward to take the limit $N \to \infty$ to get the minimal charge transfer $\mathcal{Q}_{\mathrm{B}}$ for two identical bosons. Indeed, in view of the relation~\eref{Delta_2_Delta_2_N_rel} between $\Delta_2$ and $\Delta_2^{(N)}$ and using the expression~\eref{Q_B_N_final_expr} of $\mathcal{Q}_{\mathrm{B}}^{(N)}$, we have
\begin{eqnarray}
\mathcal{Q}_{\mathrm{B}} = \lim\limits_{N \to \infty} \mathcal{Q}_{\mathrm{B}}^{(N)} = 2 \lim\limits_{N \to \infty} \lambda_{\mathrm{ring}}^{(N)} \, ,
\label{Q_B_expr}
\end{eqnarray}
and thus, since the limit in the right-hand side is the maximal backflow in the single-particle-in-a-ring case~\cite{Gou21}, we find that
\begin{eqnarray}
\mathcal{Q}_{\mathrm{B}} = - 2 c_{\mathrm{ring}} \approx - 0.233632 \, ,
\label{Q_B_final_expr}
\end{eqnarray}
where $c_{\mathrm{ring}}$ is given by~\eref{c_ring_1_part}.

Therefore, we analytically showed with~\eref{Q_B_final_expr} that the maximal amount of backflow that is achievable with two identical bosons in a ring is twice the amount of backflow achievable with a single particle in a ring. Note that the maximal amount of backflow per particle is thus equal in both the single- and the two-particle cases. Furthermore, we showed that the bosonic state that maximizes backflow is the tensor product of two single-particle backflow-maximizing states.


\subsection{Fermions}\label{id_fermions_subsec}

We now determine the minimal charge transfer $\mathcal{Q}_{\mathrm{F}}$ for two identical fermions. Here again, we first construct the minimal charge transfer $\mathcal{Q}_{\mathrm{F}}^{(N)}$ for a finite $N$. It is straightforward to see that no set of expansion coefficients $\{ c_{00}, \ldots, c_{NN} \}$ can satisfy both Eq.~\eref{c_mk_for_distinguishable_particles} and Eq.~\eref{symmetry_fermions}. Therefore, to find $\mathcal{Q}_{\mathrm{F}}^{(N)}$ requires an approach different from the one we used in section~\ref{id_bosons_subsec} for bosons.

We begin by rewriting our minimization problem in a matrix form. To this end, we organize the expansion coefficients $\{ c_{mn} \}$ into the following $(N+1)^2$-dimensional column vector:
\begin{eqnarray}
	{\bm c} = \left( \begin{array}{l} c_{00}, c_{10}, \ldots, c_{N0}, c_{01}, \ldots, c_{N1}, \ldots, c_{0N}, \ldots, c_{NN} \end{array} \right)^{\mathrm{T}} \, .
\end{eqnarray}
(For the compactness of our subsequent calculations, we do not explicitly denote the dependence of the vector ${\bm c}$ on $N$.) Then, we rewrite $\Delta_2^{(N)}$, as given by~\eref{Delta2_finite_N}, in the form
\begin{eqnarray}
	\Delta_2^{(N)} = 2 {\bm c}^{\mathrm{T}} \widehat{K} {\bm c} \, ,
\label{vec_c_Delta2}
\end{eqnarray}
where $\widehat{K}$ is the following $(N+1)^2 \times (N+1)^2$ block diagonal matrix:
\begin{eqnarray}
	\widehat{K} = \left( \begin{array}{llll}
    				K & 0 & \ldots & 0 \\
					0 & K & \ldots & 0 \\
					\vdots & \vdots & \ddots & \vdots \\
					0 & 0 & \ldots & K
					\end{array} \right) \, ,
\end{eqnarray}
where $K$ is the $(N+1) \times (N+1)$ matrix whose elements are the single-particle kernel $K_{mn}$ defined by~\eref{K_mn_1_part}, that is
\begin{eqnarray}
	K = \left( \begin{array}{llll}
    				K_{00} & K_{01} & \ldots & K_{0N} \\
					K_{10} & K_{11} & \ldots & K_{1N} \\
					\vdots & \vdots & \ddots & \vdots \\
					K_{N0} & K_{N1} & \ldots & K_{NN}
					\end{array} \right) \, .
\end{eqnarray}
We also rewrite the normalization constraint~\eref{normalization_finite_N} in matrix form:
\begin{eqnarray}
	{\bm c}^{\mathrm{T}} {\bm c} = 1 \, .
	\label{vec_c_normalization}
\end{eqnarray}

We then take into account the fermionic symmetrization constraint~\eref{symmetry_fermions} in the following way. Out of the $(N+1)^2$ expansion coefficients comprising ${\bm c}$, only $N (N+1) / 2$ are independent. Organizing these independent coefficients in the column vector 
\begin{eqnarray}
	{\bm u} = \left( \begin{array}{l} c_{10}, c_{20}, \ldots, c_{N0}, c_{21}, c_{31}, \ldots, c_{N1}, \ldots, c_{N,N-1} \end{array} \right)^{\mathrm{T}} \, ,
\end{eqnarray}
we introduce a rectangular matrix $M$ such that
\begin{eqnarray}
	{\bm c} = M {\bm u} \, .
\label{M_def}
\end{eqnarray}
The size of $M$ is $(N+1)^2 \times N (N+1) / 2$. The substitution of Eq.~\eref{M_def} into Eqs.~\eref{vec_c_Delta2} and \eref{vec_c_normalization} yields for the charge transfer $\Delta_2^{(N)}$
\begin{eqnarray}
	\Delta_2^{(N)} = 2 \bm{u}^{\mathrm{T}} M^{\mathrm{T}} \widehat{K} M {\bm u}
\label{vec_u_Delta2}
\end{eqnarray}
and for the normalization constraint
\begin{eqnarray}
	\bm{u}^{\mathrm{T}} M^{\mathrm{T}} M {\bm u} = 1 \, .
\label{vec_u_normalization}
\end{eqnarray}
The quantum statistics obeyed by the two fermions is now encoded in Eqs.~\eref{vec_u_Delta2} and \eref{vec_u_normalization}.

The matrix $M$, relating ${\bm c}$ and ${\bm u}$ through Eq.~\eref{M_def}, can be straightforwardly computed algorithmically for any $N$. For example,
\begin{eqnarray}
	M = \left( \begin{array}{lll}
    	0 & 0 & 0 \\
		1 & 0 & 0 \\
		0 & 1 & 0 \\
	   -1 & 0 & 0 \\
		0 & 0 & 0 \\
		0 & 0 & 1 \\
		0 &-1 & 0 \\
		0 & 0 &-1 \\
		0 & 0 & 0
		\end{array} \right) \quad \mbox{for} \quad N=2 \, .
\end{eqnarray}
In general, it is easy to observe that 
\begin{enumerate}
	\item $M$ consists of $N (N+1) / 2$ column vectors, each of size $(N+1)^2$,
	
	\item the column vectors are mutually orthogonal, and
	
	\item the scalar product of each column vector with itself equals two.
\end{enumerate}
This observation implies that
\begin{eqnarray}
	M^{\mathrm{T}} M = 2 I \, ,
\label{M^T_M_for_fermions}
\end{eqnarray}
where $I$ is the $\frac{N (N+1)}{2} \times \frac{N (N+1)}{2}$ unit matrix. Substituting Eq.~\eref{M^T_M_for_fermions} into Eq.~\eref{vec_u_normalization} we obtain the following (simpler) version of the normalization constraint:
\begin{eqnarray}
	2 {\bm u}^{\mathrm{T}} {\bm u} = 1 \, .
\label{vec_u_normalization_simple}
\end{eqnarray}

The minimization of $\Delta_2^{(N)}$, as given now by Eq.~\eref{vec_u_Delta2}, subject to the normalization constraint~\eref{vec_u_normalization_simple} can then be performed by means of the unconstrained minimization of the functional $\Delta_2^{(N)} - 2 \lambda {\bm u}^{\mathrm{T}} {\bm u}$, with $2 \lambda$ being a Lagrange multiplier. The corresponding Euler-Lagrange equation reads
\begin{eqnarray}
	M^{\mathrm{T}} \widehat{K} M {\bm u} = \lambda {\bm u} \, .
\label{EL_eqs_for_u}
\end{eqnarray}
Therefore, we immediately conclude from~\eref{EL_eqs_for_u} that the minimal charge transfer $\mathcal{Q}_{\mathrm{F}}^{(N)}$ is determined by the smallest eigenvalue of the matrix $M^{\mathrm{T}} \widehat{K} M$:
\begin{eqnarray}
	\mathcal{Q}_{\mathrm{F}}^{(N)} = \min \lambda \, .
\end{eqnarray}
We find this smallest eigenvalue numerically.


\begin{center}
\begin{figure}[!ht]
\centering
\includegraphics[width=1.0\linewidth]{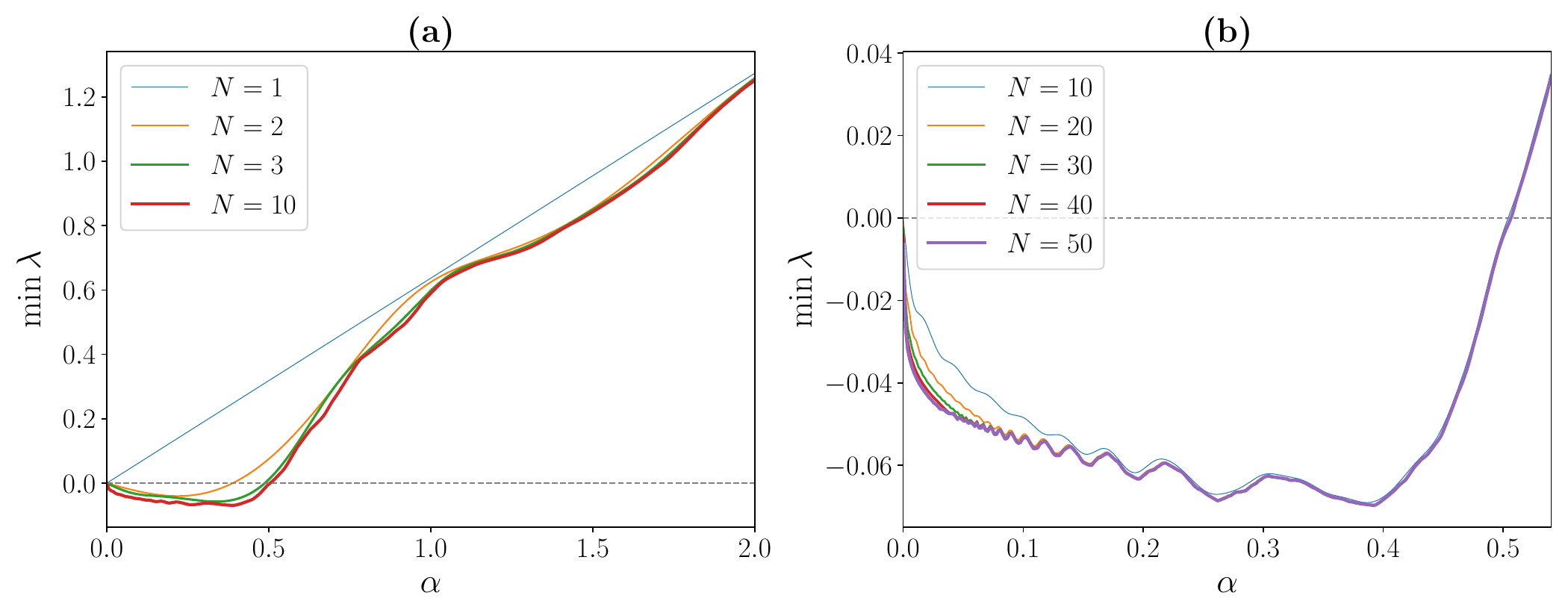}
\caption{Smallest eigenvalue $\min \lambda$ of the matrix $M^{\mathrm{T}} \widehat{K} M$ as a function of the parameter $\alpha$ for different values of $N$: panel (a) $N=1$ (blue curve), $N=2$ (orange curve), $N=3$ (green curve) and $N=10$ (red curve); panel (b) $N=10$ (blue curve), $N=20$ (orange curve), $N=30$ (green curve), $N=40$ (red curve) and $N=50$ (purple curve).}
\label{fig1}
\end{figure}
\end{center}


Since the matrix $M^{\mathrm{T}} \widehat{K} M$, and thus its smallest eigenvalue $\min \lambda$, depend on the parameter $\alpha$ [defined by~\eref{alpha_def}], we first numerically compute $\min \lambda$ as a function of $\alpha$ for different values of $N$. Our results are presented in figure~\ref{fig1}. Figure~\ref{fig1}(a) shows $\min \lambda$ with respect to $\alpha$ for $N=1$ (blue curve), $N=2$ (orange curve), $N=3$ (green curve) and $N=10$ (red curve). We can observe that $\min \lambda$ is negative only over a finite range of values of $\alpha$, namely for $\alpha \lesssim 0.5$, after which it monotonically increases with $\alpha$. We do not present the curves for $N>10$ as they are virtually indistinguishable from the $N=10$ one. Figure~\ref{fig1}(a) also illustrates the emergence of structural details as $N$ increases. These features are exhibited in more details in figure~\ref{fig1}(b), where we again plot $\min \lambda$ as a function of $\alpha$, but now focusing on the range of values of $\alpha$ where $\min \lambda$ is negative. We do this for different values of $N$: $N=10$ (blue curve), $N=20$ (orange curve), $N=30$ (green curve), $N=40$ (red curve) and $N=50$ (purple curve). We also note that $\min \lambda$ seemingly tends to a finite value in the limit $\alpha \to 0$ [that is, in view of~\eref{alpha_def}, in the limit $R \to \infty$ of an infinite radius]. A similar feature arises in the single-particle case, where the maximal backflow is seen to tend to the Bracken-Melloy bound~\eref{BM_bound} as $\alpha \to 0$~\cite{Gou21}, corresponding to the line limit.


\begin{center}
\begin{figure}[!ht]
\centering
\includegraphics[width=1.0\linewidth]{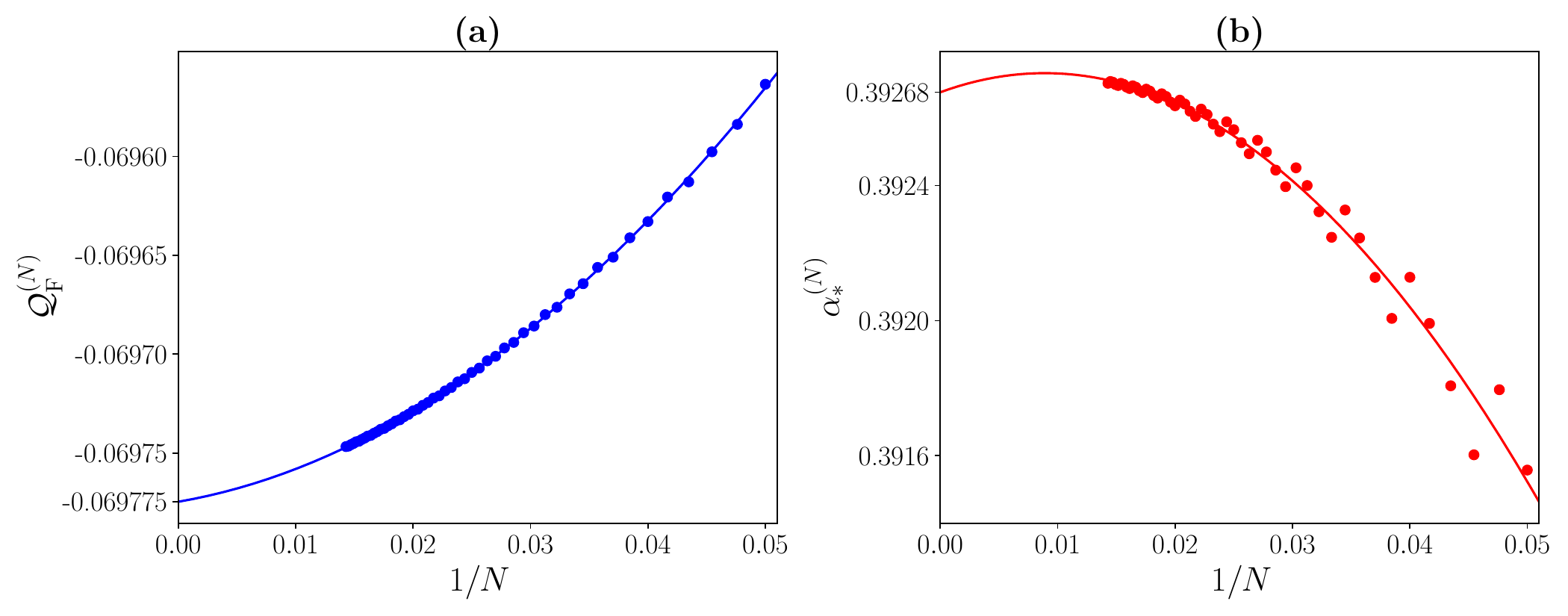}
\caption{Panel (a) Minimal charge transfer $\mathcal{Q}_{\mathrm{F}}^{(N)}$, corresponding to the global minimum of $\min \lambda$ with respect to $\alpha$, as a function of the parameter $1/N$. Panel (b) Value $\alpha_{*}^{(N)}$ of the parameter $\alpha$ at which $\min \lambda$ achieves its global minimum as a function of the parameter $1/N$. In both panels, the dots represent the computed values, while the continuous curves show the corresponding extrapolations. Further details are presented in the main text.}
\label{fig2}
\end{figure}
\end{center}


The minimal charge transfer $\mathcal{Q}_{\mathrm{F}}^{(N)}$ is then obtained by numerically evaluating the value of the global minimum of the smallest eigenvalue $\min \lambda$ with respect to the parameter $\alpha$. We evaluate $\mathcal{Q}_{\mathrm{F}}^{(N)}$ for increasing values of $N$ so as to eventually estimate the limit as $N \to \infty$, which yields the desired minimal charge transfer $\mathcal{Q}_{\mathrm{F}}$ for two identical fermions. Our results are shown in figure~\ref{fig2}. Figure~\ref{fig2}(a) shows $\mathcal{Q}_{\mathrm{F}}^{(N)}$ as a function of $1/N$: the blue dots represent the actual values of $\mathcal{Q}_{\mathrm{F}}^{(N)}$ obtained for $N=20,21, \ldots , 70$, whereas the solid blue curve is a second-degree polynomial fit to the 51 data points. This fit allows to extrapolate the limit of $\mathcal{Q}_{\mathrm{F}}^{(N)}$ as $N \to \infty$ (i.e. as $1/N \to 0$), and we find
\begin{eqnarray}
\mathcal{Q}_{\mathrm{F}} = \lim\limits_{N \to \infty} \mathcal{Q}_{\mathrm{F}}^{(N)} \approx - 0.069775 \, .
\label{Q_F_final_expr}
\end{eqnarray}

For completeness, we also show in figure~\ref{fig2}(b) the value $\alpha_{*}^{(N)}$ of the parameter $\alpha$ at which $\min \lambda$ achieves its global minimum as a function of $1/N$: the red dots correspond to the actual values of $\alpha_{*}^{(N)}$ obtained for the same 51 values of $N$ as in figure~\ref{fig2}(a), namely $N=20,21, \ldots , 70$, whereas the solid red curve is a second-degree polynomial fit, which allows to extrapolate the limit $\alpha_{*}$ of $\alpha_{*}^{(N)}$ as $N \to \infty$ and we find
\begin{eqnarray}
\alpha_{*} \equiv \lim\limits_{N \to \infty} \alpha_{*}^{(N)} \approx 0.39268 \, .
\label{alpha_star_def}
\end{eqnarray}

Finally, comparing the maximal backflow~\eref{Q_F_final_expr} in the fermionic case with the maximal backflow~\eref{Q_B_final_expr} in the bosonic case readily shows that $\mathcal{Q}_{\mathrm{F}} / \mathcal{Q}_{\mathrm{B}}$ is around $30\%$. Therefore, the nature of the particles plays a crucial role in determining the maximally achievable backflow. Specifically, using fermions appears to significantly hinder quantum backflow compared to bosons.


\section{Conclusion}\label{ccl}


\subsection{Summary}\label{summary_subsec}

In this work, we investigated the phenomenon of quantum backflow for two identical nonrelativistic particles moving freely on a circular ring. Our criterion for backflow is based on the particle-number current $J(\theta,t)$ at position $\theta$ on the ring and time $t$, defined by Eq.~\eref{J_def_2_part}: quantum backflow occurs whenever $J(\theta,t) < 0$ for a non-negative-angular-momentum state of the form~\eref{non_neg_superp_2_part}. We then constructed the charge (or number-of-particles) transfer $\Delta_2$, defined as the time integral of the current over a time interval $T > 0$, see Eq.~\eref{Delta_2_def}. Finally, we minimized $\Delta_2$ under both the normalization and the symmetrization constraints (the latter characterizing the nature, bosonic or fermionic, of the particles). This allowed us to quantify the impact of quantum statistics on the magnitude of quantum backflow in a ring.

We analytically showed that, for two identical bosons, the maximally achievable amount of backflow is exactly twice the maximal amount of backflow in the single-particle case, see Eq.~\eref{Q_B_final_expr}. Furthermore, we showed that the bosonic backflow-maximizing state is the tensor product of two single-particle backflow-maximizing states. We then numerically determined that, for two identical fermions, the maximally achievable amount of backflow is considerably smaller than in the single-particle case, see Eq.~\eref{Q_F_final_expr}. This led us to conclude that employing fermions significantly hinders quantum backflow compared to bosons. Therefore, this suggests that a prospective experimental observation of quantum backflow should be easier to perform with a bosonic system than with a fermionic one.


\subsection{Discussion}\label{discussion_subsec}

Our results can be qualitatively understood through the phenomena of boson bunching and fermion antibunching. Boson bunching refers to the tendency of bosons to occupy the same single-particle state. In the present context, the two bosons can thus accumulate in the same single-particle backflow-maximizing state. This is consistent with our finding that the maximal two-particle backflow for bosons equals twice the single-particle bound~\eref{c_ring_1_part}.

In contrast, fermion antibunching refers to the fact that fermions must obey the Pauli exclusion principle, which prevents them from occupying the same single-particle state. This has a direct consequence on quantum backflow here: the fermionic backflow-maximizing state must thus be an entangled superposition of two distinct single-particle states, each a priori deviating from the state that maximizes backflow in the single-particle case. Since, in the single-particle scenario, even a minor deviation from the backflow-maximizing state significantly reduces the magnitude of the effect, it is natural to expect that the entangled fermionic state yields a much smaller amount of backflow compared to its bosonic counterpart.

This qualitative picture also explains why, in the case of fermions, quantum backflow cannot occur for large values of the parameter $\alpha$, as can be seen in figure~\ref{fig1}(a). Indeed, in view of the definition~\eref{alpha_def} of $\alpha$, increasing $\alpha$ is equivalent to decreasing the radius $R$ of the ring. As $R$ decreases, the two fermions get spatially closer and closer to each other. As a result, the effects of Pauli's exclusion principle become more and more pronounced as $R$ decreases, hence driving the individual fermion states further away from the single-particle backflow-maximizing state.


\subsection{Outlook}\label{outlook_subsec}

A natural follow-up question is how to extend our results to systems with more than two particles. In particular, here we showed that the maximal backflow for two identical bosons in a ring is exactly twice the single-particle maximum. If we now take $N$ identical bosons and consider the state that is the tensor product of $N$ single-particle backflow-maximizing states, then following the line of reasoning of section~\ref{id_bosons_subsec} leads us to conclude that the charge transfer for this state is simply $N$ times $c_{\mathrm{ring}}$. Interestingly, at first glance, this conclusion appears to differ from the results reported in~\cite{Bar20} for $N$ identical bosons on a line, where it has been shown that quantum backflow vanishes as $N \to \infty$. Therefore, it would now be interesting to properly understand the precise connection between the formulation of quantum backflow constructed here by means of the particle current and the formulation discussed in~\cite{Bar20}, which is built on the probability of finding at least one particle at negative positions on the line. This question is of practical importance, since both of these formulations require different physical platforms to be implemented in practice: for instance, the probability used in~\cite{Bar20} cannot be defined for a periodic system such as a ring, but it can potentially be easier to measure than the particle current in a time-of-flight experiment. It is thus crucial to understand which formulation and physical setup of many-particle quantum backflow is the most adapted to the long-awaited experimental observation of the elusive quantum backflow phenomenon.


\section*{Data availability statement}

All data that support the findings of this study are included within the article (and any supplementary files).




\appendix


\section{Real coefficients \texorpdfstring{$c_{mn}$}{c}}\label{real_coef_app}

Here we show that in view of minimizing the quantity~\eref{Delta_2_expr} we can, without loss of generality, assume from the beginning that the coefficients $c_{mn}$ are real. To this end, we write $c_{mn}$ in the form
\begin{eqnarray}
c_{mn} = a_{mn} + i b_{mn} \, ,
\label{c_mn_complex_gen}
\end{eqnarray}
where $a_{mn}, b_{mn} \in \mathbb{R}$.

First, substituting~\eref{c_mn_complex_gen} into~\eref{Delta_2_expr} hence yields
\begin{eqnarray}
\fl \Delta_2 = 2 \sum_{m,n,k=0}^{\infty} \left( a_{mk} K_{mn} a_{nk} + b_{mk} K_{mn} b_{nk} \right) + 2 i \sum_{m,n,k=0}^{\infty} \left( a_{mk} K_{mn} b_{nk} - a_{nk} K_{mn} b_{mk} \right) \, .
\label{Delta_2_temp_1}
\end{eqnarray}
We can now note that making the change of indices $m,n \rightarrow n,m$ allows to write
\begin{eqnarray}
\sum_{m,n,k=0}^{\infty} a_{nk} K_{mn} b_{mk} = \sum_{m,n,k=0}^{\infty} a_{mk} K_{nm} b_{nk} \, ,
\end{eqnarray}
that is since in view of~\eref{K_mn_sym} we have $K_{nm} = K_{mn}$
\begin{eqnarray}
\sum_{m,n,k=0}^{\infty} a_{nk} K_{mn} b_{mk} = \sum_{m,n,k=0}^{\infty} a_{mk} K_{mn} b_{nk} \, ,
\end{eqnarray}
so that we get for~\eref{Delta_2_temp_1}
\begin{eqnarray}
\Delta_2 = 2 \sum_{m,n,k=0}^{\infty} \left( a_{mk} K_{mn} a_{nk} + b_{mk} K_{mn} b_{nk} \right) \, .
\label{Delta_2_temp_2}
\end{eqnarray}
Then, substituting~\eref{c_mn_complex_gen} into the normalization and symmetrization constraints~\eref{normal_superp_2_part} and~\eref{sym_postulate_c_coef} yields
\begin{eqnarray}
\sum_{m,n=0}^{\infty} \left( a_{mn}^2 + b_{mn}^2 \right) = 1
\label{norm_constraint_real}
\end{eqnarray}
and
\begin{eqnarray}
a_{n m} + i b_{nm} = \sigma a_{m n} + i \sigma b_{m n} \, ,
\end{eqnarray}
that is, equating the real and imaginary parts on both sides,
\begin{eqnarray}
a_{n m} = \sigma a_{m n} \qquad \mbox{and} \qquad b_{n m} = \sigma b_{m n} \, .
\label{sym_constraint_real}
\end{eqnarray}

Therefore, to minimize~\eref{Delta_2_expr} with respect to $c_{mn}$ under the two constraints~\eref{normal_superp_2_part} and~\eref{sym_postulate_c_coef} is equivalent to minimizing~\eref{Delta_2_temp_2} with respect to $a_{mn}$ and $b_{mn}$ under the two constraints~\eref{norm_constraint_real} and~\eref{sym_constraint_real}. We now first rewrite the quantity~\eref{Delta_2_temp_2} in the form
\begin{eqnarray}
\Delta_2 = \Delta_{2a} + \Delta_{2b} \, ,
\label{Delta_2_sum_def}
\end{eqnarray}
where we defined
\begin{eqnarray}
\Delta_{2a} = 2 \sum_{m,n,k=0}^{\infty} a_{mk} K_{mn} a_{nk}
\label{Delta_2_1_def}
\end{eqnarray}
and
\begin{eqnarray}
\Delta_{2b} = 2 \sum_{m,n,k=0}^{\infty} b_{mk} K_{mn} b_{nk} \, .
\label{Delta_2_2_def}
\end{eqnarray}
Then, we rewrite the normalization constraint~\eref{norm_constraint_real} in the form
\begin{eqnarray}
\sum_{m,n=0}^{\infty} a_{mn}^2 = \xi_1 \; \; , \; \sum_{m,n=0}^{\infty} b_{mn}^2 = \xi_2 \; \; , \; \xi_1 + \xi_2 = 1 \, .
\label{norm_constraint_real_alt}
\end{eqnarray}
Therefore, noting on~\eref{Delta_2_1_def}-\eref{Delta_2_2_def} that $\Delta_{2a}$ only involves the coefficients $a_{mn}$ and that $\Delta_{2b}$ only involves the coefficients $b_{mn}$, to minimize~\eref{Delta_2_sum_def} is equivalent to minimizing separately $\Delta_{2a}$ under the constraints
\begin{eqnarray}
\sum_{m,n=0}^{\infty} a_{mn}^2 = \xi_1 \qquad \mbox{and} \qquad a_{n m} = \sigma a_{m n}
\label{constraints_Delta_2_1}
\end{eqnarray}
and $\Delta_{2b}$ under the constraints
\begin{eqnarray}
\sum_{m,n=0}^{\infty} b_{mn}^2 = \xi_2 \qquad \mbox{and} \qquad b_{n m} = \sigma b_{m n} \, ,
\label{constraints_Delta_2_2}
\end{eqnarray}
and under the additional constraint that
\begin{eqnarray}
\xi_1 + \xi_2 = 1 \, .
\label{xi_1_xi_2_constr}
\end{eqnarray}

Now, we introduce the coefficients $\alpha_{mn}$ and $\beta_{mn}$ defined by
\begin{eqnarray}
\alpha_{mn} \equiv \frac{a_{mn}}{\sqrt{\xi_1}} \qquad \mbox{and} \qquad \beta_{mn} \equiv \frac{b_{mn}}{\sqrt{\xi_2}} \, . 
\label{alpha_beta_def}
\end{eqnarray}
This allows us to rewrite $\Delta_{2a}$ and $\Delta_{2b}$ as
\begin{eqnarray}
\Delta_{2a} = \xi_1 \bar{\Delta}_{2a} \qquad \mbox{and} \qquad \Delta_{2b} = \xi_2 \bar{\Delta}_{2b} \, ,
\label{alt_Delta_2_1_2}
\end{eqnarray}
where we defined
\begin{eqnarray}
\bar{\Delta}_{2a} = 2 \sum_{m,n,k=0}^{\infty} \alpha_{mk} K_{mn} \alpha_{nk}
\label{bar_Delta_2_1_def}
\end{eqnarray}
and
\begin{eqnarray}
\bar{\Delta}_{2b} = 2 \sum_{m,n,k=0}^{\infty} \beta_{mk} K_{mn} \beta_{nk} \, .
\label{bar_Delta_2_2_def}
\end{eqnarray}
Furthermore, substituting~\eref{alpha_beta_def} into~\eref{constraints_Delta_2_1} yields
\begin{eqnarray}
\sum_{m,n=0}^{\infty} \alpha_{mn}^2 = 1 \qquad \mbox{and} \qquad \alpha_{n m} = \sigma \alpha_{m n} \, ,
\label{alpha_constraints}
\end{eqnarray}
whereas substituting~\eref{alpha_beta_def} into~\eref{constraints_Delta_2_2} yields
\begin{eqnarray}
\sum_{m,n=0}^{\infty} \beta_{mn}^2 = 1 \qquad \mbox{and} \qquad \beta_{n m} = \sigma \beta_{m n} \, .
\label{beta_constraints}
\end{eqnarray}
Therefore, to minimize $\Delta_{2a}$ with respect to $a_{mn}$ under the constraints~\eref{constraints_Delta_2_1} is equivalent to minimizing $\bar{\Delta}_{2a}$, as given by~\eref{bar_Delta_2_1_def}, with respect to $\alpha_{mn}$ under the constraints~\eref{alpha_constraints}. Furthermore, to minimize $\Delta_{2b}$ with respect to $b_{mn}$ under the constraints~\eref{constraints_Delta_2_2} is equivalent to minimizing $\bar{\Delta}_{2b}$, as given by~\eref{bar_Delta_2_2_def}, with respect to $\beta_{mn}$ under the constraints~\eref{beta_constraints}. Comparing~\eref{bar_Delta_2_1_def} and~\eref{alpha_constraints} with~\eref{bar_Delta_2_2_def} and~\eref{beta_constraints} readily shows that the latter two constrained minimization problems are identical, so that we must have
\begin{eqnarray}
\inf\limits_{\alpha} \bar{\Delta}_{2a} = \inf\limits_{\beta} \bar{\Delta}_{2b} \equiv \bar{\Delta}_{\mathrm{inf}} \, .
\label{identical_minima}
\end{eqnarray}

Now, since combining~\eref{Delta_2_sum_def} with~\eref{alt_Delta_2_1_2} yields
\begin{eqnarray}
\Delta_2 = \xi_1 \bar{\Delta}_{2a} + \xi_2 \bar{\Delta}_{2b} \, ,
\end{eqnarray}
to minimize $\Delta_2$ with respect to $c_{mn}$ hence yields
\begin{eqnarray}
\inf\limits_{c} \Delta_2 = \xi_1 \inf\limits_{\alpha} \bar{\Delta}_{2a} + \xi_2 \inf\limits_{\beta} \bar{\Delta}_{2b} \, ,
\end{eqnarray}
that is in view of~\eref{identical_minima}
\begin{eqnarray}
\inf\limits_{c} \Delta_2 = \left( \xi_1 + \xi_2 \right) \bar{\Delta}_{\mathrm{inf}} \, ,
\end{eqnarray}
and thus in view of~\eref{xi_1_xi_2_constr} 
\begin{eqnarray}
\inf\limits_{c} \Delta_2 = \bar{\Delta}_{\mathrm{inf}} \, .
\label{inf_Delta_2}
\end{eqnarray}

This result hence shows that the problem of minimizing
\begin{eqnarray}
\Delta_2 = 2 \sum_{m,n,k=0}^{\infty} c_{mk}^* K_{mn} c_{nk}
\label{Delta_2_gen_expr_app}
\end{eqnarray}
with respect to the complex coefficients $c_{mn}$ under the two constraints
\begin{eqnarray}
\sum_{m,n=0}^{\infty} |c_{m n}|^2 = 1 \qquad \mbox{and} \qquad c_{n m} = \sigma c_{m n}
\label{constraints_def_app}
\end{eqnarray}
is \textit{equivalent} to the problem of minimizing
\begin{eqnarray}
\bar{\Delta}_2 = 2 \sum_{m,n,k=0}^{\infty} \gamma_{mk} K_{mn} \gamma_{nk}
\label{bar_Delta_2_gen_expr_app}
\end{eqnarray}
with respect to the real coefficients $\gamma_{mn}$ under the two constraints
\begin{eqnarray}
\sum_{m,n=0}^{\infty} \gamma_{m n}^2 = 1 \qquad \mbox{and} \qquad \gamma_{n m} = \sigma \gamma_{m n} \, .
\label{bar_constraints_def_app}
\end{eqnarray}
Therefore, comparing~\eref{Delta_2_gen_expr_app}-\eref{constraints_def_app} and~\eref{bar_Delta_2_gen_expr_app}-\eref{bar_constraints_def_app} indeed shows that the coefficients $c_{mn}$ can, without any loss of generality, be assumed to be real for our constrained minimization problem.


\section*{References}

\bibliographystyle{unsrt}
\bibliography{QB_id_part_ring}


\end{document}